\documentclass{article}
\usepackage[utf8]{inputenc}
\usepackage{cite}
\usepackage{amsmath}
\usepackage{graphicx}

\title{Detecting inspiratory holds in bedside waveforms using a probabilistic modeling approach}
\author{Anouk van Diepen}
\date{January 2020}

\begin{document}

\maketitle

\section{Abstract}
In this paper we propose an algorithm for finding inspiratory holds in bedside waveforms using a probabilistic modeling approach. Inspiratory holds contain useful information about the properties of the patient. Finding inspiratory holds in bedside waveforms is a problem the author ran into when working on research with flow and pressure measurements from mechanical ventilators. In this paper, we propose a simple solution for detecting the inspiratory holds. 

\section{Introduction}
\begin{figure}[t]
   \centering
 \includegraphics[width=0.8\textwidth]{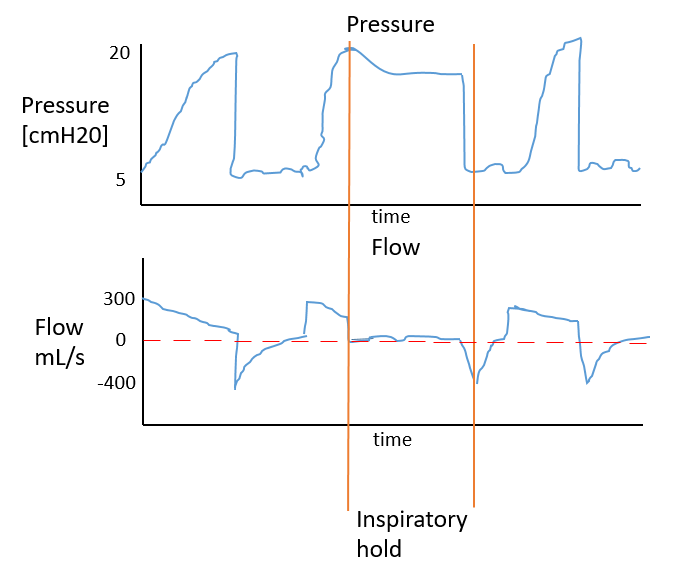}
\caption{Example of a pressure waveform and flow waveform with an inspiratory hold.}\label{fig:insp_hold}
\end{figure}

Mechanical ventilation is a basic life saving intervention on the ICU. However, ventilation can be quite dangerous and can lead to permanent lung damage if the patient is not ventilated correctly \cite{slutsky_ventilator-induced_2013}. How a patient responds to mechanical ventilation depends for a large extent on the resistance and compliance of the patient's own respiratory system. A trick to measure these parameters is called the inspiratory hold manoeuvre \cite{singer_basic_2009}. During the inspiratory hold manoeuvre, the airflow to the patient is minimized and it is possible to measure the resistance and compliance. These inspiratory holds are often made by hand and can only be done for a very short amount of time, due to the limitation in airflow.
It might be useful to look back into the measurements and to find the inspiratory holds. However, this might be very laborious due to the amount of data that is being collected. In order to assist doctors and to encourage research on mechanical ventilation, we propose an algorithm based on a probabilistic modeling approach.
Usually during mechanical ventilation, three waveforms are measured, the airflow, the volume and the pressure. During an inspiratory hold, the flow to the patient is zero and the pressure is high (see Figure \ref{fig:insp_hold}). This is a unique property of the inspiratory hold and makes it therefore very suitable for probabilistic modeling.

\section{Model specification}
From a probabilistic modeling point of view, two situations can be defined. At a time instance there is an inspiratory hold (model 1 or $M_1$) or there is no inspiratory hold (model 2 or $M_2$). To test whether the data $D$ favors $M_1$ or $M_2$, we use Bayes theorem:\\

\begin{equation}
    \frac{p(M_1|D)}{p(M_2|D)} = \frac{p(D|M_1)p(M_1)}{p(D|M_2)p(M_2)}
\end{equation}
\\
We call the term $p(D|M_x)$ the model evidence, which shows the preference of the data for a specific model, and $p(M_x)$ the prior distribution which specifies our prior preference for a certain model \cite{bishop_pattern_2006}. It is clear that the model specification is an important factor when this term is calculated. Because inspiratory holds have a distinct pattern compared to the rest of the data (no flow and a high pressure), it is not difficult to come up with a suitable model. For simplicity, we start with assuming that the flow and pressure are independent:

\begin{equation}
    p(D|M_1) = p(flow|M_1)p(pressure|M_1)
\end{equation}

To simplify the model even further, we assume that the data points in time are also independent:

\begin{equation}
    p(D|M_1) = \prod_{t=1}^{T}p(flow_t|M_1)p(pressure_t|M_1)
\end{equation}

Where $flow_t$ and $pressure_t$ are the time samples of the flow and pressure respectively at time $t$.\\

Due to the nature of the measurements, we know that the data has white noise with fixed mean and fixed variance during the inspiratory holds. We can therefore define the probability distributions over the data points in the following manner:

\begin{align}
p(flow_t|M_1) &= \mathcal{N}(flow_t|\mu_f,\sigma_f^2)\\
p(pressure_t|M_1) &= \mathcal{N}(pressure_t|\mu_p, \sigma_p^2)
\end{align}

For simplicity, and because of the knowledge about the inspiratory holds, the author chooses to set the following values for the mean and variances:

\begin{align*}
    \mu_f = 0 \\
    \sigma_f^2 = 1 \\
    \mu_p = 15 \\
    \sigma_p^2 = 1 \\
\end{align*}

For a more complicated model, there could also be chosen to learn these parameters from the data or to specify extra probability distributions over these parameters. However, in order to do this it might be helpful to have a training set, which is not available in our case.\\

The definition of our second model evidence is chosen in the following way:

\begin{equation}
    P(D|M_2) = 1-p(D|M_1)
\end{equation}

We know a priori that $M_2$ is much more likely than $M_1$, because there are only a few inspiratory holds in our data set. We can therefore choose fixed values for $p(M_1)$ and $p(M_2)$. However, because these values are constant, the following statement is valid:

\begin{equation}
    \frac{p(M_1|D)}{p(M_2|D)} \propto \frac{p(D|M_1)}{p(D|M_2)}
\end{equation}

It does not matter which constant values we choose for the prior distributions.\\

Based on all the above information, we choose the following function to be our scoring function:

\begin{equation}
    f(t) = \frac{\mathcal{N}(flow_t|\mu_f,\sigma_f^2)\mathcal{N}(pressure_t|\mu_p, \sigma_p^2)}{1-\mathcal{N}(flow_t|\mu_f,\sigma_f^2)\mathcal{N}(pressure_t|\mu_p, \sigma_p^2)}\label{eq:scoring}
\end{equation}

\section{Experimental validation}
\begin{figure}[t]
   \centering
 \includegraphics[width=0.8\textwidth]{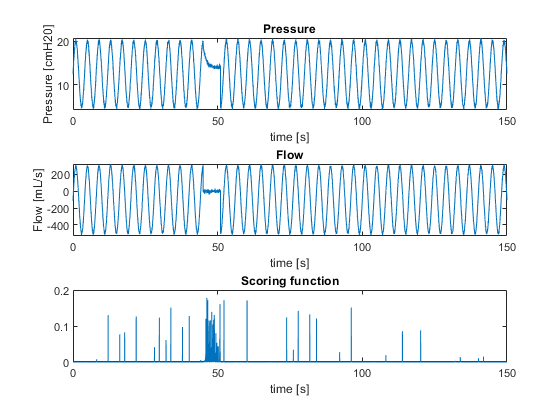}
\caption{Example of a pressure waveform and flow waveform with an inspiratory hold around 45 seconds, and the result of evaluating the scoring function (Equation 8) on these waveforms.} \label{fig:results_of_insp}
\end{figure}

The algorithm is evaluated on a real data set. However, due to privacy reasons, the author has decided not to disclose the results of the real data set. In order to illustrate how the algorithm works, we show the results on a mock waveform that has the same properties as the real waveforms. To produce the mock waveform and to evaluate the model evidence, Matlab 2019b is used. The simulations for the pressure, flow, and evaluation of the scoring function are shown in Figure \ref{fig:results_of_insp}. During the inspiratory hold, the scoring function is high, indicating a high model evidence for $M_1$. 

\section{Conclusion}
In this paper we proposed an easy solution to the problem of detecting inspiratory holds in flow waveforms and pressure waveforms using probabilistic models. 
The advantages of this approach are that the data does not need to be filtered, the model is extendable to expiratory holds, and it is possible to make the model more complex.
The current model specification is one of the simplest models that is possible. It is easy to come up with more complex models that might perform better (for example, by taking time dependence into account or taking the dependence between the pressure and the flow into account). However, for our current research, the model specification as proposed in this paper is sufficient.

\section{Acknowledgements}
Special thanks to Ashley de Bie and the Catharina Hospital in Eindhoven, for collecting the real data, sharing the data, taking the time to explain inspiratory holds, and inspiring me to make this.

\bibliographystyle{unsrt}
\bibliography{mybib}{}

\begin{thebibliography}{1}

\bibitem{slutsky_ventilator-induced_2013}
Arthur~S. Slutsky and V.~Marco Ranieri.
\newblock Ventilator-induced lung injury.
\newblock {\em The New England Journal of Medicine}, 369(22):2126--2136,
  November 2013.

\bibitem{singer_basic_2009}
Benjamin~D. Singer and Thomas~C. Corbridge.
\newblock Basic invasive mechanical ventilation:.
\newblock 102(12):1238--1245.

\bibitem{bishop_pattern_2006}
Christopher~M. Bishop.
\newblock {\em Pattern recognition and machine learning}.
\newblock Information science and statistics. Springer, New York, 2006.

\end{thebibliography}

\end{document}